\documentclass{article}
\usepackage[]{amsmath,amssymb}
\usepackage{graphics,epsfig}

\begin{document}
\date{}
\title{Universal terms for the entanglement entropy in 2+1 dimensions}
\author{H. Casini\footnote{e-mail: casini@cab.cnea.gov.ar}\,\, 
and
M. Huerta\footnote{e-mail: huerta@cabtep2.cnea.gov.ar} \\
{\sl Centro At\'omico Bariloche,
8400-S.C. de Bariloche, R\'{\i}o Negro, Argentina}}
\maketitle
\begin{abstract}
We show that the entanglement entropy and alpha entropies corresponding to spatial polygonal sets in $(2+1)$ dimensions contain a term which scales logarithmically with the cutoff. Its coefficient is a universal quantity consisting in a sum of contributions from the individual vertices. For a free scalar field this contribution is given by the trace anomaly in a three dimensional space with conical singularities located on the boundary of a plane angular sector. We find its analytic expression as a function of the angle. This is given in terms of the solution of a set of non linear ordinary differential equations. For general free fields, we also find the small-angle limit of the logarithmic coefficient, which is related to the two dimensional entropic c-functions. The calculation involves a reduction to a two dimensional problem, and as a byproduct, we obtain the trace of the Green function for a massive scalar field in a sphere where boundary conditions are specified on a segment of a great circle. This also gives the exact expression for the entropies for a scalar field in a two dimensional de Sitter space.  
\end{abstract}

\vspace{1.cm}



%\noindent arXiv: hep-th/0606256 

\section{Introduction}
Suppose we are interested in the physics of events localized in a region $V$ of the space. The Hilbert space of states can be decomposed accordingly as a tensor product $
 {\cal H}={\cal H}_{V}\otimes {\cal H}_{-V}$
 of the spaces of the states localized in $V$ and  in the complementary region $-V$. Take now the vacuum $\left| \Psi \right>$ as a global state  of the system, with density matrix $
\rho_0= \left| \Psi \right> \left< \Psi \right|$. 
 The state $\rho_V$ relevant to the algebra of operators acting on ${\cal H}_{V}$
  follows from the partial trace of $\rho_0$ over the complementary Hilbert space ${\cal H}_{-V}$. This gives the local reduced density matrix 
\begin{equation}
 \rho_{V}=\textrm{tr}_{{\cal H}_{-V}}\left| \Psi \right> \left< \Psi \right|\,.
\end{equation}
The global state $\rho_0$ is generally entangled in the bipartite system ${\cal H}_{V}\otimes {\cal H}_{-V}$ and in consequence this matrix is mixed.  
The corresponding entropy
 \begin{equation}
 S(V)=-\textrm{tr}(\rho_V\log \rho_V)\,
\end{equation}
is usually called entanglement or geometric entropy. 
 
The entanglement entropy is one of the most prominent candidates to explain the intriguing entropy of the black holes \cite{bombelli}. However, in this proposal the role of quantum gravity is fundamental to produce a finite entropy, and the whole subject is still controversial. 
On the other hand, this and other measures of entanglement have also been extensively studied in condensed matter and low dimensional systems, partially motivated by advances in quantum information theory and the density matrix renormalization group method.  As a result it was uncovered that a variety of phenomena such as quantum phase transitions  have an interesting  correlate in the entanglement properties of fundamental states \cite{phase}.  
From the point of view of quantum field theory (QFT) the function $S(V)$ can be considered as a non local variable with interesting non perturbative properties \cite{hg}, which can be defined for any theory disregarding the field content. In this context, we can mention among the applications the description of topological order \cite{topological}, and the renormalization group irreversibility in two dimensions \cite{cteor,rg}. 

In the continuum limit described by a quantum field theory the entanglement entropy is divergent due to the presence of an unbounded number of local degrees of freedom. The divergent terms must be proportional to quantities which are local and extensive on the boundary of $V$.
This can be seen as a consequence of the local nature of the ultraviolet divergences, and that the boundary is shared between $V$ and $-V$ which have the same entropy (for any pure global state)
\begin{equation}
S(V)=S(-V)\,.
\end{equation}
On a technical level this characteristic of the divergent terms is due the fact that the entanglement entropy is the variation of the euclidean free energy with respect to conical singularities located on the boundary of $V$ \cite{tech}. Thus, in $d$ spatial dimensions, we expect to have an expansion of the form
 \begin{equation}
 S(V)=g_{d-1}[\partial V] \,\epsilon^{-(d-1)} + g_{d-2}[\partial V]\,\epsilon^{-(d-2)} +...+ g_0[\partial V]\,\log (\epsilon \Lambda)+ S_0(V)\,,   \label{div}
 \end{equation}
 where $S_0(V)$ is a finite part, $\epsilon$ is a short distance cutoff, and the $g_i$ are local and extensive  functions on the boundary $\partial V$, which are homogeneous of degree $i$ (in a different context a similar expansion was used in \cite{ryu}). The leading divergent term coefficient  $g_{d-1}[\partial V]$ is proportional to the $d-1$ power of the size of $V$. This was noted since the earliest papers on the subject \cite{bombelli,sred} and is usually referred to as the area law for the entanglement entropy. However, strictly speaking, $g_{d-1}$  depend on the regularization procedure and it is not proportional to the area if this later is not rotational invariant.
 For example, if we draw a square of side $L$ on a two dimensional lattice,  we would have $g_1(\partial V)\sim C_1\frac{L}{\epsilon}$, where $\epsilon$ is the lattice spacing, but where the dimensionless constant $C_1$ strongly depends on the relative angle between the square sides and the lattice symmetry axes.
 Moreover, all the terms $g_i$ with $i\ge 1$ are not physical within QFT since they are not related to continuum quantities. 
 
On the contrary, the dimensionless coefficient $g_0(V)$ of the logarithmic term is expected to be universal. 
Logarithmic divergent terms in the entropy have been previously found in four dimensional black hole space-times \cite{ss}. They are present generically in even dimensions for sets with smooth curved boundaries. This follows from the heat kernel expansion for conical manifolds with smooth singularity surface \cite{conical}. 

In this work we show that there is also a logarithmic term in three dimensions for sets $V$ with non-smooth boundary. 
In particular, we consider the case of spatial polygonal sets  (in a different scenario, a logarithmic contribution to the entanglement entropy was also reported in \cite{diferent}). Since $g_0(V)$ is dimensionless, extensive and local on the boundary we conclude that for $V$ a polygon it must be of the form
\begin{equation}
g_0(V)=\sum_{v_i} s(x_i)\,,
\end{equation}       
where the sum is over all vertices $v_i$ and  $x_i$ is the vertex angle. On general grounds one also expects point-like vertex induced logarithmic  terms in any dimensions.

All these considerations apply to the alpha-entropies (or R\'enyi entropies) as well
\begin{equation}
S_\alpha(V)=\frac{1}{1-\alpha}\log(\textrm{tr}\rho_V^\alpha),\,\,\,\,\,\, S(V)=\lim_{\alpha\rightarrow 1} S_\alpha (V) \,.\label{cuatro}
\end{equation} 
 These measures of information behave very much like the entanglement entropy, as discussed in several papers \cite{fermion,cahu}. For integer $\alpha=n$ they are also more suitable for explicit calculation, since the traces $\textrm{tr}\rho_V^n$ involved can be represented by a functional integral on an n-sheeted space with conical singularities located at the boundary of $V$ \cite{tech}. The entanglement entropy follows from $S_n$ by analytical continuation. 
Thus, in 2+1 dimensional theories the alpha-entropies contain a logarithmic term 
\begin{equation}
\left.S_n\right|_{\log}=\sum_{v_i} s_n(x_i) \log(\epsilon \Lambda)\,, \label{cinco}
\end{equation} 
analogous to the one in (\ref{div}) for the entropy. We also have $s(x)=\lim_{\alpha\rightarrow 1} s_\alpha(x)$. 
In eq. (\ref{cinco}) $\Lambda$ is a parameter with the dimensions of an energy, depending on $V$ and on the particular  theory. For a massless field it is the inverse of any typical dimension $R$ of $V$ and when the mass dominates, $MR\gg 1$, it can be taken  as $\Lambda=M$. We find the analytic expression of $s_n(x)$ for a free scalar field, and check our results by numerical simulations on a two dimensional lattice. 

The outline of the paper is the following. In Section 2, we obtain the analytic expression for $\textrm{tr}\rho_V^n$ for the local density matrix associated to a plane angular sector for a massive scalar field in $2+1$ dimensions. The calculation is reduced to the one of the trace of the Green function on a two dimensional sphere, where boundary conditions are specified on a segment of a great circle.  
We use the method introduced in \cite{cahu} to obtain this trace exactly.
 Finally, we identify the logarithmic coefficients $s_n(x)$ of (\ref{cinco}). We also compare them  with the results given by simulations in a square lattice finding a perfect accord.
In Section 3, we find the small angle limit of $s(x)$ for free fields and relate it to the two dimensional entropic c-functions. Finally, in Section 4 we present our conclusions. We have also included in the Appendix A a detailed derivation of the formulas concerning the Green function in the sphere with a cut presented in Section 2.1, and in Appendix B, the method we have used to compute numerically the geometric entropy in a two dimensional lattice.   

\section{Vertex induced logarithmic terms in the entropy}

To compute the coefficients $s_n(x)$ of the logarithmic term in (\ref{cinco}) for a scalar field we choose for convenience to study the entropy associated to a set $V$ given by a plane angular sector of angle $x$. 

The traces $\textrm{tr}\rho_V^{\alpha }$ involved in (\ref{cuatro}), with $\alpha=n\in
{\mathbb Z}$, can be represented by a functional integral on an $n$-sheeted three-dimensional Euclidean space with conical singularities
located at the boundary of the set $V$ \cite{tech,fermion}. To be explicit, calling $u$ and $v$ to the upper and lower faces of the plane angular sector, the replicated space is obtained considering $n$ copies of the three-space cut along the angular sector, and sewing together the upper side of the cut $u^{k}$ with
the lower one $v^{k+1}$, for the different copies $ k=1,...,n$, and where the copy $n+1$ coincides with the first one.
The trace of $\rho^{n}$ is then given by the functional integral
$Z[n]$ for the field in this manifold, 
\begin{equation}
\textrm{tr}\rho^{n}= \frac{Z[n]}{Z[1]^{n}}\,. 
\label{dd}
\end{equation}

Then, following \cite{fermion, cahu} we consider a free massive complex scalar field on this manifold, and map the problem to an equivalent one in which we deal with $n$ decoupled and multivalued free complex scalar fields. First we arrange the values of the field in the different copies in a single vector field living in a three dimensional space,
\begin{equation}
\vec{\Phi}=\left(\begin{array}{c} 
\phi _{1}(\vec{x}) \\ 
\vdots \\  
\phi _{n}(\vec{x}) \end{array}
\right) \,,
\end{equation}
where $\phi _{l}(\vec{x})$ is the field on the $l^{\textrm{th}}$ copy. 
Note that in this way the space is simply connected but the singularities at the boundaries of $V$ are still there since the vector $\vec{\Phi}$ is not singled valued.  In fact, crossing the plane angular sector from above (side $u$) or from below (side $v$), the field gets multiplied by a matrix $T$ or $T^{-1}$ respectively. Here
\begin{equation}
\begin{array}{c}
T=\left(
\begin{array}{lllll}
0 & 1 &  &  &  \\
& 0 & 1 &  &  \\
&  & . & . &  \\
&  &  & 0 & 1 \\
1 &  &  &  & 0
\end{array}
\right)
\end{array}\,, 
\end{equation}
which has eigenvalues\footnote{There is an important difference with the free fermion case studied in \cite{fermion}, where the eigenvalues of $T$ are $e^{i\frac{k}{n}2\pi }$, with $k=-(n-1)/2\,,...,\,(n-1)/2$. This is because in the fermionic case it is $T^n=-1$.} $e^{i\frac{k}{n}2\pi }$, with
$k=0\,,...,\,(n-1)$. 
Then, changing basis by a unitary transformation in the replica space,
we can diagonalize $T$, and the problem is reduced to $n$ decoupled
fields $\tilde{\phi}_{k}$ living on a single three dimensional space. These fields are multivalued and defined on the euclidean three dimensional space with boundary conditions imposed on the two dimensional set $V$ given by
\begin{equation}
\tilde{\phi}^u_k(\vec{r})=e^{i\frac{2 \pi k}{n}}\tilde{\phi}^v_k(\vec{r})\,\,\,\,\,\,\,\,\,\,k=0,\,\,...\,n-1, \,\,\,\,\,\,\, \, \vec{r}\in V\,.
\label{bc}
\end{equation}
Here $\tilde{\phi}^{u}_k$ and $\tilde{\phi}^{v}_k$ are the limits of the field as the variable approaches $V$ from each of its two opposite sides in three dimensions. 
In this formulation we have
\begin{equation}
\log(\textrm{tr}\rho_V^n)=\sum_{k=0}^{n-1}\textrm{log}Z[k/n]\,,\label{sumz}
\end{equation}
where $Z[a]$ is the partition function corresponding to a field which acquires a phase $e^{i2\pi a}$ when the variable crosses $V$. We note that the coefficient of the logarithmic divergent term in $\log Z[k/n]$ corresponds to the integrated trace anomaly  $\int dr^3 T^\mu_\mu(\vec{r})$, or equivalently the $a_3=c_{3/2}/(4\pi)^{3/2}$ coefficient of the heat kernel expansion \cite{vassi}, for a massless scalar in three dimensions with the boundary conditions (\ref{bc}).

Our calculation of $Z[a]$ is based on the relation between the free energy and the Green function for a free massive scalar
\begin{equation}
\partial_{M^2}\textrm{log}Z[a]=-\int dr^3 G_a(\vec{r},\vec{r})\,.
\label{green}
\end{equation}
Here $G_a(\vec{r}_1,\vec{r}_2)$ is the Green function for a complex scalar of mass $M$ in three dimensions subject to the boundary conditions (\ref{bc}). To be explicit, we have
\begin{eqnarray}
(-\Delta_{\vec{r}_1}+M^2) \,G_a(\vec{r}_1,\vec{r}_2)&=&\delta(\vec{r}_1-\vec{r}_2)\,,\\
\lim_{\varepsilon\rightarrow 0^+} G_a(\vec{r}_1+\varepsilon \hat{\eta},\vec{r}_2)&=&e^{i 2 \pi a}  \,\lim_{\varepsilon\rightarrow 0^+} G_a(\vec{r}_1-\varepsilon \hat{\eta},\vec{r}_2)\,, \hspace{1cm} r_1\in V\, ,\label{cator}
\end{eqnarray} 
 where  $\hat{\eta}$ is orthogonal to the plane of $V$.
 
The Laplacian and the boundary conditions allow the separation of angular and radial equations in polar coordinates. Using standard methods we arrive at the expression (see for example the similar calculation in \cite{cc})
\begin{equation}
G_a(\vec{r}_1,\vec{r}_2)=\sum_\nu \int d\lambda \frac{\lambda}{\lambda^2+M^2}\psi_{\nu}( \theta_1 , \varphi_1)\psi^*_{\nu}(\theta_2 , \varphi_2)\frac{J_{\frac{1}{2}+\nu}(\lambda r_1)J_{\frac{1}{2}+\nu}(\lambda r_2)}{\sqrt{r_1 r_2}}\,,
\label{ga}
\end{equation}
where $J$ is the standard Bessel function. Here the sum is over the normalized eigenvectors $\psi_{\nu}(\theta , \varphi)$ of the angular equation 
\begin{equation}
\Delta_{\Omega}\psi_{\nu}=-\nu(\nu+1)\psi_{\nu}\,,
\end{equation}
where $\Delta_{\Omega}$ is the Laplacian on the sphere with domain given by the functions satisfiyng the boundary conditions inherited from (\ref{cator}). Specifically we can choose the cut on the equatorial plane 
\begin{equation}
\lim_{\varepsilon\rightarrow 0^+} \psi_{\nu}(\pi/2+\varepsilon , \varphi)= e^{i 2 \pi a}  \lim_{\varepsilon\rightarrow 0^+} \psi_{\nu}(\pi/2-\varepsilon , \varphi)\,,  \hspace{1.3cm} \varphi \in [-x/2,x/2]\,.\label{boun}
\end{equation}
The functions $\psi_{\nu}$ are the spherical harmonics if $a=0$, otherwise they are known to satisfy Lam\'e differential equations \cite{lame}. Also, the values of $\nu$ are not integer if $a\neq 0$. The precise expressions for $\psi_{\nu}$ and $\nu$ will not be relevant in what follows.  

Taking the trace $\int dr^3 G_a(\vec{r},\vec{r})$ in eq. (\ref{ga}) gives
\begin{equation}
\partial_{M^2}\log Z[a]=-\frac{1}{2M^2}\sum_{\nu}(\nu+1/2)=-\frac{1}{2M^2}\textrm{tr}\sqrt{-\Delta_{\Omega}+\frac{1}{4}}\,.
\label{trsqrt}
\end{equation}
Though this expression is divergent, the piece we are interested in, which is the one dependent on the angle $x$, is finite. 
 
To proceed, we find convenient to express the trace of the square root of the operator in (\ref{trsqrt}) in terms of the corresponding Green function. This is done by using the expression for the powers of an elliptic operator $O$ in terms of the resolvent given in \cite{seeley},  
\begin{equation}
O^\gamma=\frac{i}{2 \pi}\int_{\Gamma}\lambda^\gamma(O-\lambda)^{-1}d\lambda\,,
\end{equation}
where $\Gamma$ is a curve depending on the particular operator $O$. In the present case $\Gamma$ begins at infinity, pass  along the negative real axes on the upper complex plane, encircles the origin and goes back to infinity on the lower half of the complex plane along the negative real axes.  This gives for the trace
\begin{equation}
\textrm{tr}  \sqrt{-\Delta_{\Omega}+\frac{1}{4}}=-\frac{1}{\pi}\,\int_0^{\infty}\lambda^{\frac{1}{2}}\,\,\textrm{tr}\frac{1}{\Delta_{\Omega}-\frac{1}{4}-\lambda}\, d\lambda 
\,.
\label{tr}
\end{equation}
 
\subsection{Green function on a sphere with a cut}
The problem is then reduced to the calculation of the trace of the two dimensional Green function on a sphere with a cut of angle $x$, where the boundary conditions (\ref{boun}) are imposed. In a previous paper \cite{cahu} we have solved the analogous problem on the plane with boundary conditions imposed on an interval,  using a generalization of a method originally developed in \cite{myers}. It essentially consists in exploiting the symmetries of the Helmholtz equation even in the presence of symmetry breaking boundary conditions by analyzing the behavior of the Green function at the singular points.
 
Following the recipe of \cite{cahu} step by step (with the additional algebraic complications corresponding to the spherical case) we find the analytic expression for the trace of the Green function as a solution of a system of ordinary differential equations. The details of the derivation are given in the appendix A. Explicitly we find   
\begin{equation}
\textrm{tr}\frac{1}{\Delta_{\Omega}-m^2}=-8\pi (1-a)a\int_{x}^{\pi}H_a(y)dy\,.\label{quince}
\end{equation}
The function $H_a(x)$ is the solution of the following set of ordinary non linear differential equations (we omit the subscript $a$ and the dependence on $x$ of the variables for notational convenience)
\begin{eqnarray} 
H' &=& - \frac{m}{2}\,\left( b\,B_2 +c\, B_1 + 2\,u\,B_{12} \right)\label{hprima}\,,\\
X_1' &=& - m\,\left( b\,B_{12}+ u\,B_1 \right)\label{x1prima} \,,\\
X_2' &=& - m\,\left( c\,B_{12} + u\,B_2 \right) \label{x2prima}\,,\\
c' &=&- 2\,m\,\beta_2 \,u\, \csc(x)\,\sin(x/2) - c\,(1 - a)\,\csc(x)\,(1 + \cos(x))\label{cprima}\,, \\
b' &=& -2\,m\,\beta_1\,u\,\csc(x)\,\sin(x/2) - b\,a\,\csc(x)\,(1 + \cos(x)) \label{bprima} \,,\\
u'&=&-\frac{m}{2}\,\sec(x/2)\,(b\,\beta_2+ c\,\beta_1) + \frac{1}{2}\,u\,\tan \left(x/2 \right)\label{uprima} \,,
\end{eqnarray}
where $B_1$, $B_2$, $B_{12}$, $\beta_1$, $\beta_2$ are functions of $x$ given in terms of $H$, $X_1$, $X_2$, $c$, $b$, and $u$ by the following set of algebraic equations
\begin{eqnarray}
\frac{\cos(x/2)}{8 
\pi a(1 - a)}&=&\sin(x/2)\,H - m \left(\beta_1\, X_2 + \beta_2\, X_1\right) + 2\,m\,\cos(x/2)\, u\, B_{12}\,,\label{29} \\
\frac{\sin(x/2)}{8 \pi a(1 - a)}&=&-\cos(x/2)\,H - m\, \tan(x/2)\left(\beta_1 X_2 + \beta_2 X_1 \right) + 
  m\,\sin(x/2)(b B_2 + c B_1) \,,\label{30}\\
0&=& -m \sin(x/2)(c X_1
 -b X_2) + m \tan(x/2)(\beta_2 B_1- \beta_1 B_2)+(1-2a)\cos(x/2) B_{12}\,,\label{31}\\
0&=&-4a(a - 1) - m^2(4 - 8\beta_1\beta_2 + b c + 3u^2) \nonumber\\&&\hspace{2cm}- 
      4\cos(x)\left(a(a - 1) + m^2(u^2 + 1)\right) 
      + m^2\cos(2x)(b\,c - u^2) \label{4a}\,,\\
0&=&(2a - 1)u\cos(x/2)+m \tan(x/2)(\beta_1 c - b \beta_2)\,.\label{minuscula}
\end{eqnarray}
The boundary conditions at $x\rightarrow\pi$ are
\begin{eqnarray}
H(\pi)&=&0\,,\label{hache}\\
X_1(\pi)&=&\frac{ \Gamma(-a) \left( \cosh \left( \frac{\pi \mu}{2} \right) \textrm{Im} \left[ \psi \left( \frac{1}{2} + a + \frac{i\mu}{2} \right) \right] - \frac{\pi}{2} \sinh \left( \frac{\pi\mu}{2} \right)   \right)}{2^{2a}\mu  \left( \cos \left( 2 a \pi\right)+\cosh (\pi \mu)\right) \Gamma (1+a) \left| \Gamma \left( \frac{1}{2}-a+\frac{i\mu}{2}\right) \right|^2 }\,,\label{xx1}\\
X_2(\pi)&=&X_1(\pi)\left|_{\,a\rightarrow (1-a)} \right. \,,\label{xx2} \\
u(\pi)&=&0\,, \label{upi}\\
b	(\pi)&=&\frac{ 2^{1-2a} a (1-a) \left| \Gamma\left( \frac{1}{2} +a+\frac{i\mu}{2}\right)\right|^2}{m\Gamma^2 (1+a)}\,,\label{treintaytres}\\
c(\pi)&=&b(\pi)\left|_{\,a\rightarrow (1-a)}\right.  \,,\label{treintaydos}
\end{eqnarray}
where $\mu=\sqrt{4\,m^2-1}$ and $\psi$ is the digamma function. The meaning of the extra variables $B_1$, $B_2$, $B_{12}$, $X_1$, $X_2$, $u$, $b$, $c$, $\beta_1$ and $\beta_2$ is the same as in \cite{cahu} and is given in appendix A. The trace in (\ref{quince}) is regularized such that it vanishes when $x=\pi$, where there is no vertex point and no logarithmic term is present in the entropies. 

In  \cite{doyon} the partition function for a Dirac fermion on the Poincar\'e disk with boundary conditions analogous to (\ref{boun}) imposed on a geodesic segment  has been written in terms of a solution of the Painlev\'e VI differential equation. Though we were not able to find an explicit relation, it is likely that our results for $H(x)$ could have also an expression in terms of solutions of these type of equations. 

\begin{figure} [tbp]
\centering
\leavevmode
\epsfysize=6cm
\bigskip
\epsfbox{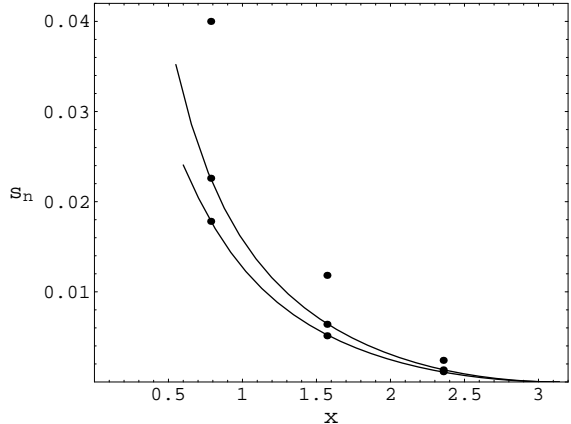}
\caption{The solid curves are the functions $s_2(x)$ (top) and $s_3(x)$ (bottom). The points are obtained by numerical simulations in a lattice, and correspond to, from top to bottom, $s=s_1$, $s_2$ and $s_3$ evaluated for the angles $x=\pi / 4$, $\pi /2$ and $3/4 \,\pi$.}
\end{figure}

\subsection{The coefficient of the logarithmic term}
 Gathering all the results together, using eqs. (\ref{cuatro}), (\ref{cinco}) with $\Lambda =M$, (\ref{sumz}), (\ref{trsqrt}), (\ref{tr}), and (\ref{quince}), we arrive at the following result for a real scalar 
\begin{equation}
s_n(x)= \sum_{k=1}^{n-1}\frac{8\,k\,(n-k)}{n^2\,(n-1)}\int_{1/2}^{\infty}dm\,m\,(m^2-1/4)^{\frac{1}{2}}\int_{x}^{\pi}dy \,H_\frac{k}{n}(y,m)\,,
\label{final}
\end{equation}
where we have made explicit the $m$ dependence of $H$.  
This equation, together with (\ref{hprima}-\ref{treintaytres}) gives our final expression for the coefficient of the logarithmic term in $S_n$ for a real scalar field (half the complex scalar result), which is in position to be evaluated by solving the ordinary differential equations numerically. 

The functions $s_n(x)$ satisfy  $s_n(x)=s_n(2 \pi-x)$, which is a consequence of the symmetry in the entropies $S_n(V)=S_n(-V)$ due to the purity of the vacuum state. In the figure (1) we have plotted $s_2(x)$ and $s_3(x)$ for $x\in[0,\pi]$. The values of $s_1=s$, $s_2$ and $s_3$ for $x=\pi / 4$, $\pi /2$ and $3/4 \,\pi$ obtained by lattice simulations are also plotted. They  show a perfect accord (less than one percent error) with the analytical results. These particular values of the angle are the ones for which the coefficient can be calculated in absolute terms with very small error on a square lattice of limited size (in the present case it was $100\times 200$ points). The numerical methods consist of evaluating the entropy for a massless real scalar  (see Appendix B and  \cite{cahu,peschel}) for a given shape (square, triangle, etc.) and different overall size $\lambda$, and then fitting the result as $S_n=C_0+C_1\, \lambda+C_{-1}\, \lambda^{-1} +C_{-2} \,\lambda^{-2}- s_n \log(\lambda)$. It is also possible to evaluate  very accurately $s_n$ for specific combinations of angles using rectangular triangles. In this way we have computed in the lattice $s_n (y)+s_n (\pi/2-y)$ with $y=\arctan(p/q)$, where $p$ and $q$ are small integers. We also obtain in this case a perfect accord with the analytical results.  We have also checked that $s_n$ does not depend on the orientation of the polygon with respect to the lattice simmetry axes.   

\section{The small angle limit  and the two dimensional entropic c-functions}
From  (\ref{final}) and the differential equations for $H$, it is possible to derive expansions for $s_n(x)$ and $s(x)$ for small $x$ or $\pi-x$. In particular, in the small $x$ limit the Green function on the cut sphere is related to the corresponding one for the flat space problem with boundary conditions imposed on an interval. Taking into account the results of \cite{cahu}, which deal with this later problem, we can show directly from the differential equations that for small $x$ 
\begin{equation}
s(x)\sim \frac{\int_0^\infty dt \, c_S(t)}{\pi x}\,,
\label{35}
\end{equation}
where $c_S(t)$ is the one dimensional entropic c-function for a free scalar field \cite{cteor,fermion,cahu}. The c-function for a theory in $1+1$ dimensions is defined as 
\begin{equation}
c(r)=r\frac{dS(r)}{dr}\,,
\end{equation}
where $S(r)$ is the entanglement entropy corresponding to an interval of length $r$ in $1+1$ dimensions. For free fields  $c(r)\equiv c(t)$ where $t=m r$, and $m$ is the field mass. 

We can obtain the formula (\ref{35}) (and the analog one which is valid for free fermions) with a less technical derivation. This also sheds light on the origin and the necessity of the logarithmic term. In a previous paper we have shown that the entropy corresponding to a spatial rectangle in $2+1$ dimensions with a short side $L$ and long side $R$, $R/L\gg 1$, contains the universal term (included in the finite term $S_0$ in eq.(\ref{div})) \cite{cahu} 
\begin{equation}
S_{\textrm{univ}}\simeq -k R/L\,, \label{tete}
\end{equation}
where $k$ is a dimensionless function of $L$ and the renormalized parameters of the theory. For a free massless theory it is 
\begin{equation}
k=\frac{1}{\pi}\int_0^\infty dt \, (n_S c_S(t)+n_F c_F(t))\,, 
\end{equation}
where $c_S$ and $c_F$ are the one dimensional entropic c-functions for a real scalar and a majorana fermion, and $n_S$ and $n_F$ are the multiplicity of the boson and fermion degree of freedom. 
This gives $k\simeq 0.039$ for a real scalar and $k\simeq 0.072$ for a $2+1$ dimensional Dirac fermion. Now, we may approximately decompose a plane angular sector with small angle $x$ as a union of thin and long rectangles with bigger size as we move further from the angle vertex. Thus, there is a term in the entropy for the angular sector which is due to the sum of the terms (\ref{tete}) for the rectangles (the term (\ref{tete}) is extensive in the direction in which lie the different rectangles). In the limit of an infinite partition this gives the desired result
\begin{equation}
\left.S\right|_{\log}\sim -k\int_\epsilon \frac{dR}{\tan(x)R} \sim \frac{\int_0^\infty dt \, c(t)}{\pi \,x} \log(\epsilon)\,.
\label{38}
\end{equation}

\begin{figure} [tb]
\centering
\leavevmode
\epsfysize=4cm
\bigskip
\epsfbox{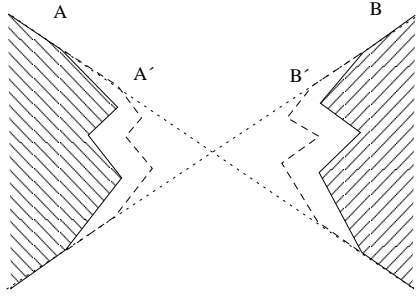}
\caption{The dashed sets $A$ and $B$ are anti-starshaped, in the sense that any ray from the origin has no intersection with them or the intersection is a half line which does not contain the origin. A contraction of the space transforms them into $A^\prime$ and $B^\prime$. In the limit of an infinite contraction they are mapped to two angular sectors with common vertex.}
\end{figure}

\section{Final remarks}
We have shown that there is a logarithmic divergent term with universal coefficient for the entanglement entropy corresponding to spatial polygonal sets in 2+1 dimensions. We have found the analytic expression for the coefficient in the alpha entropies for integer $\alpha=n$ and a free scalar field, and the small angle limit for the corresponding coefficient on the geometric entropy for general free fields. 

Interestingly, since this paper first appeared as a preprint, a vertex induced logarithmically divergent term in three dimensions has also been found in the conjectured geometrical form of the entanglement entropy for some conformal field theories  arising in the AdS/CFT duality context \cite{ryu,ht}. The overall form of the logarithmic coefficient as a function of the angle is similar to our exact results, and the authors have also been able to show the relation between $k$ and $s$ (eqs. (\ref{tete}) and (\ref{38})) for their entropy functions, which presumably correspond  to some interacting theory.

Technically, the  calculation in this paper amounts to the one of the conformal anomaly for a free scalar on a three-dimensional manifold with a conical singularity located on the boundary of a plane angular sector. This is a non-trivial result, since there is no known general method applicable to this case involving conical singularity surfaces which are themselves non-smooth. 
We have shown that this particular problem can be mapped to the one of the calculation of the trace of the Green function for a massive scalar field on a two dimensional sphere where boundary conditions are imposed on a segment of a great circle. We have found this trace analytically by a method introduced in \cite{cahu,myers} which allows one to exploit the rotational symmetries of the sphere even in the presence of the symmetry breaking boundary conditions.  

Our results for the Green function on the cut sphere may find different uses beyond the present one. One possible application is the problem of scattering of waves by a plane angular sector in three dimensions \cite{lame}. 
Remarkably, it also gives the exact entropy functions for a spatial segment in $1+1$ dimensional de Sitter space. This space is equivalent to the surface    
\begin{equation}
\vec{r}^2-t^2=R^2                      
\end{equation}
in $2+1$ Minkowski space. The euclidean de Sitter space then corresponds to a sphere with radius $R$. The formulas corresponding to (\ref{sumz}) and (\ref{green}) in this case take us again to the Green function on the cut sphere. The result can then be readily written off,
\begin{equation}
S_n^{\textrm{de Sitter}}(r/R)=- \sum_{k=1}^{n-1} \frac{8 \pi k (n-k)}{n^2 (n-1)} \int_{(MR)^2}^\infty dm^2\,   \int_{r/R}^\pi H_{\frac{k}{n}} (y,m)\, dy  + \textrm{cons} \,,
 \end{equation}
where  $r$ is the physical size of the geodesic interval, and $M$ is the field mass. The additive constant is logarithmically divergent with the cutoff, in order to keep $S_n$ positive for small values of $r$.
 
The formulas (\ref{35}), (\ref{tete}) and (\ref{38}) show a remarkable relation between universal terms in the entropies for different dimensions. The function $c(r)=r\, dS(r)/dr$, with $S(r)$ the one dimensional entropy corresponding to an interval of length $r$, plays the role of the Zamolodchikov's c-function in the entanglement entropy c-theorem \cite{cteor}. This theorem states that there is a universal dimensionless quantity in two dimensions which is decreasing under scaling and has a well defined value at the fixed points (proportional to the Virasoro central charge)\cite{zamo}. The c-function introduced by Zamolodchikov is constructed from a correlator of stress tensor traces, and we have shown that it can be taken to be $c(r)$ as well. There has been a substantial effort to extend the c-theorem to higher dimensions, but a definitive result in this direction is still missing \cite{sofar}. 
At fixed points the function $c(r)$ is a constant given by the coefficient of the term proportional to $\log(\epsilon)$ in the two dimensional entropy. Then, one would be tempted to speculate that a running function which takes the value $s(x)$ at fixed points could be a good candidate for a c-function in three dimensions. 

On the other hand, the logarithmic term in the entropy induced by the vertices is, remarkably, the only obstacle in the following simple argument attempting to prove the c-theorem in any dimensions. 
Consider the mutual information 
\begin{equation}
I(A,B)=S(A)+S(B)-S(A\cup B) \label{mar}
\end{equation}
 between two non intersecting sets $A$ and $B$. This is dimensionless and universal (all boundary terms get subtracted), and it is also positive and increasing with the size of each set separately \cite{cteor}. Thus, if we take $A$ and $B$ as anti-starshaped (see figure (2)) $I(A,B)$ is decreasing under dilatations. However, it fails to have a well-defined value at the ultraviolet fixed point, since in that limit $A$ and $B$ go to angular sectors with a common vertex, and $I(A,B)$ has a logarithmic divergence due to the mismatch of the vertex induced terms in (\ref{mar}).     
 
\section*{Acknowledgments}
We warmly thank C.D.Fosco for very useful discussions and important input in the early phase of this project, 
and R.Trinchero for pointing us to ref. \cite{seeley}. 

\section{Appendix A: Green function on the cut sphere}
In this appendix we derive the set of non linear differential equations (\ref{hprima}-\ref{treintaytres}) which give the trace of the Green function on a sphere with a cut of angle $x$ (see figure 3), where the boundary conditions (\ref{boun}) are imposed. We will follow here the same steps as in \cite{cahu} where we have solved the analogous problem on the plane with boundary conditions imposed on a finite interval.  We use a generalization of a method originally developed in \cite{myers}.

The Green function $G(z,z^\prime)=(-\Delta_{\Omega} +m^{2})^{-1}_{z,z^\prime}$, where $z$ and $z^{\prime}$ are points on the sphere, is uniquely defined  by the following three requirements:
\bigskip

\noindent a.- It satisfies the Helmholtz equation on the sphere
\begin{equation}
\left( -\Delta_{\Omega} +m^{2}\right) G(z,z^{\prime})=\delta (
z-z^{\prime})\,, \label{equ1}
\end{equation}
where, in polar coordinates,
\begin{equation}
\Delta_{\Omega}=\frac{1}{\sin \theta}\frac{\partial}{\partial \theta}(\sin\theta\frac{\partial}{\partial \theta})+\frac{1}{\sin^2\theta}\frac{\partial^2}{\partial \varphi^2}\,.
\end{equation}
\bigskip

\noindent b.- The boundary condition is (it also holds for the Green function derivatives)
\begin{equation}
\lim_{\epsilon \rightarrow 0^{+}}G((\pi+\varepsilon,\varphi ),z^{\prime})=e^{i2\pi a }\lim_{\epsilon \rightarrow 0^{+}}G((\pi-\varepsilon,\varphi ),z^{\prime}) \hspace{1cm} \textrm{for}\,\, \varphi \in [\varphi_1,\varphi_2]\,. 
\end{equation}
 Here the points with polar coordinates $(\pi/2,\varphi)$, $\varphi \in [\varphi_1,\varphi_2]$, form the cut location. We need to specify the cut through its two endpoints with angle $\varphi_1$ and $\varphi_2$. At the end of the calculation we can set as in Section 2 $\varphi_1=-\varphi_2=x/2$. From now on we also choose $a \in [0,1]$.

\bigskip 

\noindent c.- $G(z,z^{\prime})$ is bounded everywhere (including the cut) except at $z=z^{\prime}$.
\bigskip

\noindent Several properties of the Green function can be derived from its definition. First it is Hermitian  
\begin{equation}
G(z,z^{\prime})=G(z^{\prime},z)^{*}\,,
\end{equation}
and the two reflection symmetries of the problem give 
\begin{eqnarray}
G(z,z^{\prime})&=&G(Tz,Tz^{\prime})^{*}\,,\\
G(z,z^{\prime})&=&G(Rz,Rz^{\prime})\,,
\label{sime1}
\end{eqnarray}
where $R$ and $T$ are the reflexion operations given by
\begin{eqnarray}
R\,\,(\theta,\varphi)&=&(\theta,\varphi_1+\varphi_2-\varphi)\,,\\
T\,\,(\theta,\varphi)&=&(\pi-\theta,\varphi)\,. 
\end{eqnarray}

In what follows we study the structure of singularities of some functions related to $G(z,z^\prime)$ and use repeatedly the fact that a nonsingular solution of the homogeneous Helmholtz equation must be zero due to the uniqueness theorem.  

According to the boundary conditions, near the end points of the cut the Green function must
have branch cut singularities. The requirement that
the function must remain bounded on the cut and the equation (\ref{equ1})
imply that the leading terms of $G(z,z^{\prime})$ for $(\theta,\varphi)$ near $
(\pi/2,\varphi_1)$ (and fixed $z^{\prime}$) have to be of the form
\begin{equation}
G(z,z^{\prime})\sim [(\varphi-\varphi_1)-i(\theta-\pi/2)]^{a }S_{1}(z^{\prime})+[(\varphi-\varphi_1)+i(\theta-\pi/2)]^{1-a }S_{2}(z^{\prime}) \,, \label{equ2}
\end{equation}
for some functions $S_1(z)$ and $S_2(z)$. 
We have written explicitly only the terms with powers of $[(\varphi-\varphi_1)-i(\theta-\pi/2)]$ with exponent smaller than one. These are the ones of interest in the sequel. Note that the contributions at this order must be  
analytic or anti-analytic in $z$ in
order to cancel the Laplacian term in (\ref{equ1}).

The most singular contributions to $\partial _{\varphi_{1}}G(z,z^{\prime})$
for $z\rightarrow (\pi/2,\varphi_{1})$ follow from the derivative of (\ref{equ2})  
\begin{equation}
\partial _{\varphi_{1}}G(z,z^{\prime})\sim -a [(\varphi-\varphi_1)-i(\theta-\pi/2)]^{a
-1}S_{1}(z^{\prime})+(a -1)[(\varphi-\varphi_1)-i(\theta-\pi/2)]^{-a
}S_{2}(z^{\prime})  \,.\label{deri}
\end{equation}
The function $\partial _{\varphi_{1}}G(z,z^{\prime})$ satisfies the homogeneous Helmholtz equation and the boundary conditions, and it is not singular at $
z\rightarrow z^{\prime}$. It has only one singularity located at $\varphi_1$ whose expression is given by (\ref{deri}). Thus, an adequate linear combination of this function for different values of $z^\prime$ must be a nonsingular solution of the Helmholtz equation, and therefore identically zero. 
Following the same argument as in the flat case \cite{cahu}, this leads to the fundamental relation
\begin{equation}
\partial _{\varphi_{1}}G(z,z^{\prime})=S(z)^{\dagger}AS(z^{\prime})  \label{c}\,,
\end{equation}
where we are using vectorial notation 
\begin{equation}
S(z)=\left(
\begin{array}{l}
S_1(z)  \\
 S_2(z)   
\end{array}\right)
\end{equation}
and $A$ is an hermitian matrix.

The function $S(z)$ satisfies the homogeneous Helmholtz equation with the
same boundary conditions as $G(z^{\prime},z)$ with $z^{\prime}$
fixed, but it is unbounded around $z=(\pi/2,\varphi_{1})$. In fact, it follows from (\ref{deri}) and (\ref{c}) that it has singular terms proportional to 
$[(\varphi-\varphi_1)-i(\theta-\pi/2)]^{-a }$ and $[(\varphi-\varphi_1)+i(\theta-\pi/2)]^{a -1}$. 
 However, these 
disappear if we derive with respect to $\varphi_{2}$.
On the other hand $S(z)$ behaves around $(\pi/2,\varphi_{2})$ as an ordinary wave, that
is, it vanishes proportionally to $[(\varphi-\varphi_2)-i(\theta-\pi/2)]^{a }$ and $
[(\varphi-\varphi_2)+i(\theta-\pi/2)]^{1-a }$. Then $\partial _{\varphi_{2}}S(z)$ has singular
terms around $(\pi/2,\varphi_{2})$ which are proportional to $[(\varphi-\varphi_2)-i(\theta-\pi/2)]^{a -1}$ and $ [(\varphi-\varphi_1)+i(\theta-\pi/2)]^{-a }$. 
This structure of singularities  and the relation (\ref{sime1}) gives place to 
\begin{eqnarray}
\partial _{\varphi_{2}}S(z)&=&\gamma \,S(Rz) \,, \label{a}\\
\partial _{\varphi_{1}}S(Rz) &=&-\gamma \,S(z)\,, \label{b} 
\end{eqnarray}
where the matrix $\gamma $ is a function of $\varphi_{1}-\,\varphi_{2}$.

\begin{figure} [tb]
\centering
\leavevmode
\epsfysize=5cm
\bigskip
\epsfbox{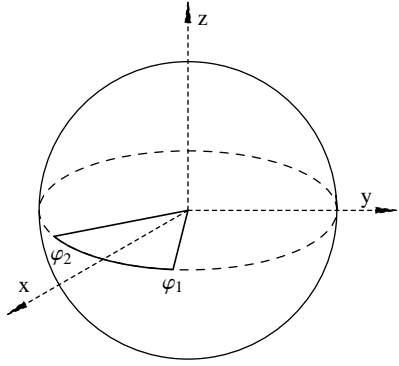}
\caption{The plane angular sector of angle $x=\varphi_1-\varphi_2$ and the sphere with a cut of angle $x$}
\end{figure}

Using (\ref{c}), (\ref{a}) and (\ref{b}) to calculate $\partial
_{\varphi_{1}}\partial _{\varphi_{2}}G(z,z^{\prime})-\partial _{\varphi_{2}}\partial
_{\varphi_{1}}G(z,z^{\prime})=0$ we get 
\begin{equation}
\gamma ^{\dagger }=A\gamma A^{-1}\,, \label{alge1}
\end{equation}
and that the matrix $A$ must be a constant, $\partial A / \partial_{\varphi_1}=\partial A/ \partial_{\varphi_2}= 0$. Thus, it can be evaluated by the knowledge of the solutions for $x=\pi$ or using the flat space  limit $x\rightarrow 0$ \cite{cahu}. It is given by
\begin{equation}
A=-4\pi (1-a )a \,\sigma _{1}\,, \label{algea} 
\end{equation}
where $\sigma _{1}$ is the Pauli matrix. 

The equation (\ref{deri}) leads to the behavior 
\begin{equation}
S(z)\sim \frac{1}{4\pi }\left(
\begin{array}{l}
\frac{1}{a }[(\varphi-\varphi_1)-i(\theta-\pi/2)]^{-a } \\
\frac{1}{(1-a )}[(\varphi-\varphi_1)+i(\theta-\pi/2)]^{a -1}
\end{array}
\right)  \label{wr}
\end{equation}
for $z$ in the vicinity of $(\pi/2,\varphi_{1})$. Using (\ref{b}) we get a similar expression for S(Rz)

\begin{equation}
S(Rz)\sim \frac{\gamma}{4\pi a(1-a)}\left(
\begin{array}{l}
\left[(\varphi -\varphi_1)-i(\theta-\pi/2)\right]^{1-a } \\ 
\left[(\varphi -\varphi_1)+i(\theta-\pi/2)\right]^{a}
\end{array}
\right) \,. \label{wrr}
\end{equation}

\subsection{Equations for $S(z)$ from rotation symmetries}
In order to use eq. (\ref{c}) to compute the trace of the Green function we need more information on $S(z)$. With this aim we exploit the symmetries of the problem as follows.  
We consider two rotations around the axes lying on the equator which cut the sphere at the points $(\theta=\pi/2, \varphi=\frac{\varphi_1+\varphi_2}{2})$ and $(\theta=\pi/2, \varphi=\frac{\varphi_1+\varphi_2}{2}+\frac{\pi}{2})$. The associated differential operators are given respectively by
\begin{eqnarray}
L_1&=&-\cot\theta\sin(\varphi-\varphi^{+})\partial\varphi+\cos(\varphi-\varphi^{+})\partial\theta \label{defl1}\,, \\
L_2&=&-\cot\theta\cos(\varphi-\varphi^{+})\partial\varphi-\sin(\varphi-\varphi^{+})\partial\theta\,. \label{defl2}
\end{eqnarray}
 We use the notation $\varphi^{\pm}=\frac{\varphi_1\pm\varphi_2}{2}$. From (\ref{defl1}) and (\ref{defl2}) we check that they satisfy the algebra of angular momentum operators
\begin{eqnarray}
\left[L_1,L_2\right]&=&\partial_{\varphi}\,,\\
\left[L_1,\partial_{\varphi}\right]&=&-L_2\,,\\
\left[L_2,\partial_{\varphi}\right]&=&L_1\,,\\
L_1^2+L_2^2+\partial^2_{\varphi}&=&\Delta_{\Omega}\,,
\end{eqnarray}
and the following relations
\begin{eqnarray}
\frac{\partial L_1}{\partial\varphi_1}&=&-\frac{1}{2}L_2\,,\\
\frac{\partial L_2}{\partial\varphi_1}&=&\frac{1}{2}L_1\,.
\end{eqnarray}
 
They commute with the Laplacian operator and
thus the functions $(L_i+L_i^{\prime})\,G(z,z^{\prime})$ with $i=1,2$ 
are non singular at $z=z^{\prime}$ and satisfy the
homogeneous Helmholtz equation and the boundary conditions. 
However, due to the $L_i G(z,z^{\prime})$ terms, they are singular at $(\pi/2,\varphi_{1})$ and $(\pi/2,\varphi_{2})$ as a function of $z$. Near $(\pi/2,\varphi_{1})$ we have from (\ref{equ2}), (\ref{wr}), (\ref{defl1}) and (\ref{defl2})
\begin{eqnarray}
L_1 G(z,z^{\prime})&\sim& \cos\varphi^{-}\left[i a [(\varphi-\varphi_1)-i(\theta-\pi/2)]^{a
-1}S_{1}(z^{\prime})-i(1-a )[(\varphi-\varphi_1)+i(\theta-\pi/2)]^{-a
}S_{2}(z^{\prime})\right] \nonumber \\
&\sim& -i\cos(\varphi^{-}) S(z^{\prime})A \sigma_3 S(Tz)\,,\\
L_2 G(z,z^{\prime})&\sim& i \sin(\varphi^{-})S(z^{\prime})A \sigma_3 S(Tz)\,.
\end{eqnarray}
This gives from the uniqueness of the solution of the homogeneous Helmholtz equation for non-singular functions 
\begin{eqnarray}
L_1 G(z,z^{\prime})+L_1^{\prime} G(z,z^{\prime})&=& -i\cos(\varphi^{-}) \left[S^{\dagger}(z) \sigma_3A S(z^{\prime})+ S^{\dagger}(Rz) \sigma_3A S(Rz^{\prime})\right]\,,\label{dL1}\\
L_2 G(z,z^{\prime})+L_2^{\prime} G(z,z^{\prime})&=& -i\sin(\varphi^{-}) \left[S^{\dagger}(z) \sigma_3A S(z^{\prime})- S^{\dagger}(Rz) \sigma_3A S(Rz^{\prime})\right]\,.\label{dL2}
\end{eqnarray}

From (\ref{wr})  we see that the most divergent terms of $L_1 S^{\dagger}(z)A- i\cos(\varphi^{-})\partial_{\varphi}S^{\dagger}(z)\sigma_3 A$ exactly cancel. But this combination should also have a contribution to order $[(\varphi-\varphi_2)-i(\theta-\pi/2)]^{a -1}$ and $ [(\varphi-\varphi_1)+i(\theta-\pi/2)]^{-a }$ for $z$ in the vicinity of $(\pi/2,\varphi_{1})$. Thus we have 
\begin{equation}
L_1 S^{\dagger}(z)A-i \cos(\varphi^{-})\partial_{\varphi}S^{\dagger}(z)\sigma_3 A=-S^{\dagger}(z)A \xi \label{chi}\,,
\end{equation}
where $\xi$ is a matrix which depends on $x=\varphi_1-\varphi_2$.
A similar argument hold for $L_2 S^{\dagger}(z)A  + i \sin(\varphi^{-} 
) $ $\partial  \varphi S^{\dagger}(z) \sigma_3  A$ giving place to 
\begin{equation}
L_2 S^{\dagger}(z)A +i \sin(\varphi^{-})\partial\varphi S^{\dagger}(z)\sigma_3 A=S^{\dagger}(z)A \xi_1 \label{chi1}\,,
\end{equation}
with $\xi_1$ depending on $x=\varphi_1-\varphi_2$.

Taking the derivative with respect to $\varphi_1$ of (\ref{dL1}) and (\ref{dL2}), and extracting the coefficients of the divergent terms, we get
\begin{eqnarray}
\label{bata}
L_1 S(z)&=&\xi S(z)-i \sin(\varphi^{-})\sigma_3 S(z)-i\cos\varphi^{-}\left[\left\{ \gamma ,\sigma _{3}\right\}S(Rz)-\sigma_3 \partial_{\varphi}S(z)\right]\,,\\
\label{tata}
L_2 S(z)&=&-\xi_1S(z)-i\cos(\varphi^{-})\sigma_3 S(z)+i\sin(\varphi^{-})\left[\left[\sigma_3,\gamma\right]S(Rz)+\sigma_3\partial_{\varphi}S(z)\right]\,.
\end{eqnarray}
We can also write the reflected equation for (\ref{bata}) and (\ref{tata})
\begin{eqnarray}
L_1 S(Rz)=\xi S(Rz)-i \sin\varphi^{-}\sigma_3 S(Rz)-i\cos\varphi^{-}\left[\left\{ \gamma ,\sigma _{3}\right\}S(z)-\sigma_3 \partial_{\varphi}S(Rz)\right]  \,,\label{bata1}\\
L_2 S(Rz)=\xi_1S(Rz)+i\cos(\varphi^{-})\sigma_3 S(Rz)-i\sin(\varphi^{-})\left[\left[\sigma_3,\gamma\right]S(Rz)-\sigma_3\partial_{\varphi}S(z)\right]\,.\label{tata1}
\end{eqnarray}
Subtracting the hermitian conjugate of the equation (\ref{bata}) and (\ref{bata1}) and  using $\left\{ A,\sigma
_{3}\right\} =0$, $\left\{ \gamma ,\sigma _{3}\right\} ^{\dagger }A+A\left\{
\gamma ,\sigma _{3}\right\} =0$, we obtain
\begin{equation}
\xi ^{\dagger }A+A\xi =-i \sin \varphi^{-} \sigma_3 A \,. \label{alge2}\\
\end{equation}
A similar relation for $\xi_1$ follows by subtracting the hermitian conjugate of (\ref{tata}) and (\ref{tata1}),
\begin{equation}
\xi_1 ^{\dagger }A+A\xi_1=i \cos \varphi^{-} \sigma_3 A \,. \label{alge2p}
\end{equation}
\subsection{Parametrization}
The expansion of $S(z)$ can be extended to the following order around the singular point $(\pi/2,\varphi_1)$ introducing a real matrix $N$
\begin{equation}
S(z)\sim \frac{1}{4\pi }\left(
\begin{array}{l}
\frac{1}{a }[(\varphi-\varphi_1)-i(\theta-\pi/2)]^{-a } \\
\frac{1}{(1-a )}[(\varphi-\varphi_1)+i(\theta-\pi/2)]^{a -1}
\end{array}
\right) +N  \left(
\begin{array}{l}
\left[(\varphi-\varphi_1)-i(\theta-\pi/2)\right]^{1-a } \\
\left[(\varphi-\varphi_1)+i(\theta-\pi/2)\right]^{a}
\end{array}
\right)\,.\label{2orden}
\end{equation}
This general expansion for $S$ inserted in (\ref{chi}) and  (\ref{chi1}) allows us to solve for some components of $N$ and obtain a relation between $\xi$ and $\xi_1$
\begin{equation}
\xi_1=\tan (\varphi^{-})\xi+i q \sec(\varphi^{-})\,,
\end{equation}
with
\begin{equation}
q=\left(
\begin{array}{ll}
a-1 &  \\
& a
\end{array}
\right)\,.
\end{equation}
It also gives additional relations which, together with the algebraic equations (\ref{alge1}), (\ref{alge2}), give the general form of $\gamma$ and $\xi$ in terms of the  
following parametrization 
\begin{equation}
\gamma =\frac{m}{2}\left(
\begin{array}{ll}
u & b \\
c & u
\end{array}
\right)
\hspace{1.5cm};\hspace{1.5cm}\xi
=\left(
\begin{array}{ll}
-i(a-1)\sin(\varphi^{-}) & im\beta _{1} \\
-im\beta _{2} & -ia\sin(\varphi^{-})
\end{array}
\right)\,,
\end{equation}
where $u$, $b$, $c$, $\beta _{1}$, and $\beta _{2}$ are real functions of $m$ and $x$.

Taking derivatives respect to $\theta$ and $\varphi$ of (\ref{bata}) and (\ref{tata}) and combining them to reconstruct the Helmholtz equation for $S(z)$, and using the expansions of $S(z)$ and $S(Rz)$, we get from the series around the singular points
 the differential equations (\ref{cprima}), (\ref{bprima}), (\ref{uprima}) for $b$, $c$ and $u$, and the algebraic equations (\ref{4a}) and  (\ref{minuscula}) which determine $\beta_1$ and $\beta_2$.  
\subsection{Integrated quantities}
The equation (\ref{c}) gives for the trace of the Green function
\begin{equation}
\frac{d}{dx}\int d\Omega\, G(z,z)=\int d\Omega\,  S^{\dagger
}A\;S=-8\pi a \left( 1-a \right) H \,,\nonumber 
\end{equation}
where 
\begin{equation}
H(x) =\int d\Omega \,S_{1}^{*}(z)S_{2}(z) \,, \label{hh}
\end{equation}
and $d\Omega=d\theta d\varphi \sin(\theta)$.

To find this quantity, we use the information obtained in 5.1 and 5.2. We first define the following auxiliary integrals 
\begin{eqnarray}
B_{1}(x) &=&\int d\Omega \,S_{1}^{*}(z)S_{1}(Rz)\,, \\
B_{2}(x) &=&\int d\Omega\,S_{2}^{*}(z)S_{2}(Rz) \,,\\
B_{12}(x) &=&\int d\Omega\,S_{2}^{*}(z)S_{1}(Rz) \,,\\
X_{1}(x) &=&\int d\Omega\,S_{1}^{*}(z)S_{1}(z) \,,\label{x1}\\
X_{2}(x) &=&\int d\Omega\,S_{2}^{*}(z)S_{2}(z)\,.\label{x2}
\end{eqnarray}
These are convergent. They are also real since the relation (\ref{sime1}) implies that $S_{i}^{*}(z)=S_{i}(Tz)$. 

The equations for these quantities are obtained basically combining conveniently the components of (\ref{bata}) and (\ref{tata}) multiplied by the components of $S(z)$ and $S(Rz)$, and integrating on the sphere. In this way the combination $\left(L_1\,S_1(z)\right)S_2^{*}(z)+S_1(z)\left(L_1\,S_2(z)\right)^{*}$ integrated on the sphere gives place to the equation (\ref{29}). 
Similarly, from the combination $\left(L_2S_1(z)\right)S_2^{*}(z)+S_1(z)\left(L_2S_2(z)\right)^{*}$ we get equation (\ref{30}) and from $\left(L_2S_2(Rz)^{*}\right)S_1(z)+S_2(Rz)^{*}\left(L_2S_1(z)\right)^{*}$ we get equation (\ref{31}).

Differential equations for the integrated variables follow by taking the derivative of (\ref{hh}), (\ref{x1}) and (\ref{x2}) with respect to $\varphi_{2}$. 
These correspond to (\ref{hprima}), (\ref{x1prima}) and (\ref{x2prima}) of Section  2.1. 

In the flat space analog of the present calculation it is possible to give a closed algebraic expression of $H$ in terms of $u$ \cite{cahu}. This simplification seems to be lost in the sphere. 

\subsection{Boundary conditions at $x=\pi$}
The homogeneous Helmholtz equation on the cut sphere for a function $f(z)$ can be solved exactly by separation of variables when $x=\pi$. The solution has the general form
\begin{equation}
f(z)=\sum_{n=-\infty}^{\infty} f_n(\theta)e^{i\varphi(a+n)}\,,
\end{equation}
where we have put now the end points of the cut at the poles $\theta=0$ and $\theta=\pi$ of the polar coordinates.  Here $f_n$ is given by
\begin{equation}
f_n(\theta)=C_1 P_{\frac{1}{2}(-1+i\mu)}^{a+n}(w)+C_2 Q_{\frac{1}{2}(-1+i\mu)}^{a+n}(w)\,,
\end{equation}
with $Q$ and $P$ the standard Legendre functions, $w=\cos\theta$ and $\mu=\sqrt{4m^2-1}$.
Evaluating the limits $w\rightarrow1$ and $w\rightarrow-1$ and comparing with the expansion (\ref{wr}) for $S(z)$  we get for $x=\pi$
\begin{eqnarray}
S_1&=&\frac{2^{-a/2}}{4\pi a}\frac{1}{2^{\frac{a}{2}-1}\cos(a\pi-i\frac{\pi}{2}\mu)\csc (a\pi)\Gamma(a)} \left[\frac{\pi}{2}\cot(a\pi)P_{\frac{1}{2}(-1+i\mu)}^{a}(w) \label{s1pi}
-Q_{\frac{1}{2}(-1+i\mu)}^{a}(w)\right]e^{i\varphi a} \,,  \\
S_2&=&\frac{2^{(a-1)/2}}{4\pi (1-a)}\frac{1}{2^{\frac{-(a+1)}{2}}\cos(a\pi+i\frac{\pi}{2}\mu)
\csc (a\pi)\Gamma(1-a)}\left[\frac{\pi}{2}\cot((1-a)\pi)P_{\frac{1}{2}(-1+i\mu)}^{a-1}(w)\right.\nonumber \\ &-&\left.Q_{\frac{1}{2}(-1+i\mu)}^{a-1}(w)\right]e^{i\varphi (a-1)} \,. \label{s2pi}
\end{eqnarray}
Using these expressions, the eqs. (\ref{a}) and (\ref{b}), and the expansion of $S(Rz)$ we find $u(\pi)$, $b(\pi)$ and $c(\pi)$ given in (\ref{upi}), (\ref{treintaytres}) and (\ref{treintaydos}).
The results (\ref{hache}), (\ref{xx1}) and (\ref{xx2}) for $H(\pi)$, $X_1(\pi)$ and $X_2(\pi)$ follow directly from (\ref{hh}), (\ref{x1}) and  (\ref{x2}), and the explicit form of $S_1$  and $S_2$, where we have used some formulae for the integrals of a product of two Legendre functions given in \cite{higher}.    

\section{Appendix B: numerical entropy in a two dimensional lattice}
We use the method presented in \cite{peschel} to give an expression for $\rho _{V}$  in
terms of correlators for free bosonic discrete systems. 
Take a free Hamiltonian for bosonic degrees of freedom with the form  
\begin{equation}
H=\frac{1}{2}\sum_{\vec{r}} \pi _{\vec{r}}^{2}+\frac{1}{2}\sum_{\vec{r}\vec{r}^\prime}\phi _{\vec{r}}M_{\vec{r}\vec{r}^\prime}\,\phi
_{\vec{r}^\prime}\,\,,
\end{equation}
where $\phi _{\vec{r}}$ and $\pi _{\vec{r}}$ obey the canonical commutation relations $
[\phi _{\vec{r}},\pi _{\vec{r}^\prime}]=i\delta _{\vec{r}\vec{r}^\prime}$, $M$ is a Hermitian positive
definite matrix, and the sums are over the lattice sites $\vec{r}$. The vacuum (ground state) correlators are given by 
\begin{eqnarray}
X_{\vec{r}\vec{r}^\prime} &=&\left\langle \phi _{\vec{r}}\phi _{\vec{r}^\prime}\right\rangle =\frac{1}{2}(M^{-
\frac{1}{2}})_{\vec{r}\vec{r}^\prime}\,,  \label{x} \\
P_{\vec{r}\vec{r}^\prime} &=&\left\langle \pi _{\vec{r}}\pi _{\vec{r}^\prime}\right\rangle =
\frac{1}{2}(M^{\frac{1
}{2}})_{\vec{r}\vec{r}^\prime}\,.  \label{p}
\end{eqnarray}
Let $X_{\vec{r}\vec{r}^\prime}^{V}$ and $P_{\vec{r}\vec{r}^\prime}^{V}\,$ be the correlator matrices restricted to
the region $V$, that is $\vec{r},\vec{r}^\prime\in V$. 
The entropies  can be evaluated as 
\begin{eqnarray}
S&=&\sum_k \left((\nu _{k}+1/2)\log (\nu _{k}+1/2)-(\nu
_{k}-1/2)\log (\nu _{k}-1/2) \right)\,,  \label{for}  \\
S_n&=&\frac{1}{n-1}\sum_k \log\left( \left( \nu_k+1/2 \right)^n - \left( \nu_k-1/2 \right)^n\right)\,,
\end{eqnarray}
where $\nu_k$ are the eigenvalues of  $\sqrt{X^{V}.P^{V}}$.
Thus, to compute the entropy numerically we need to diagonalize the matrix $
X^{V}.P^{V}$.

We took the lattice Hamiltonian for a real massless scalar in three dimensions as   
\begin{equation}
{\cal H}=\frac{1}{2}\sum_{n,m=-\infty}^{\infty}( \pi _{nm}^{2}+(\phi _{n+1,m}-\phi_{n,m})^{2}+(\phi _{n,m+1}-\phi_{n,m})^{2} )\,.
\end{equation}
We have set the lattice spacing to one. 
The correlators (\ref{x}) and (\ref{p}) are  
\begin{eqnarray}
\langle \phi_{00}\phi_{ij}\rangle &=&\frac{1}{8\pi^2}\int_{-\pi}^{\pi} \int_{-\pi}^{\pi}dxdy\frac{\cos(x(i-1))\cos(y(j-1))}{\sqrt{2
(1-\cos(x))+2
(1-\cos(y))}}\,, \\
\langle \pi_{00}\pi_{ij}\rangle &=& \frac{1}{8\pi^2}\int_{-\pi}^{\pi} \int_{-\pi}^{\pi}dxdy \cos(x(i-1))\cos(y(j-1))}{\sqrt{2
(1-\cos(x))+2
(1-\cos(y))}\,.
\end{eqnarray}
It is relevant to the accuracy of the entropy calculation to evaluate the correlators with enough precision. This can be very time-consuming. We have found it is much faster to evaluate one of the two integrals in the correlators analytically (in terms of polynomials times elliptic functions) using a program for analytic mathematical manipulations, and then doing the last integral numerically.

\end{document}